\documentclass[12pt]{iopart}

\usepackage{iopams}
\usepackage{graphicx}
\usepackage{subfig}
\usepackage{color}
\usepackage{ulem}
\usepackage{color}
\usepackage{ulem}
\usepackage{hyperref}

\begin{document}
	
	\title{Triggering of tearing instability by impurity radiation through resistive interchange reversal in a tokamak}
	
	
	\author{Shiyong Zeng}
	\address{Department of Plasma Physics and Fusion Engineering, University of Science and Technology of China, Hefei, Anhui 230026, China}
	
	\author{Ping Zhu}
	\address{International Joint Research Laboratory of Magnetic Confinement Fusion and Plasma Physics, State Key Laboratory of Advanced Electromagnetic Engineering and Technology, School of Electrical and Electronic Engineering, Huazhong University of Science and Technology, Wuhan, Hubei 430074, China}
	\address{Department of Engineering Physics, University of Wisconsin-Madison, Madison, Wisconsin 53706, USA}
	\ead{zhup@hust.edu.cn}
	
	\author{Ruijie Zhou}
	\address{Institute of Plasma Physics, Chinese Academy of Sciences, Hefei, Anhui 230031, China}

	\author{Dominique Frank Escande}
	\address{Aix-Marseille University, CNRS, PIIM, UMR 7345 Marseille, France}
	
	
	\title[The triggering of TM by radiation through resistive interchange reversal]{}
	
	\newpage
	\begin{abstract}
		Recent MHD simulations find that the reversal of the local resistive interchange parameter $D_R$ from negative to positive due to impurity radiation cooling is able to trigger the resistive tearing mode growth in a tokamak above a threshold in impurity level. A layer of perturbed Pfirsch-Schl\"{u}ter current density and resistivity are also induced by the impurity radiation, which further govern the tearing mode growth and saturation in the nonlinear stage. The impurity threshold and the tearing mode growth strongly depend on the parallel thermal conductivity, and such a dependence derives from the impact on $D_R$ of the fast parallel thermal equilibration along the helical magnetic field lines.
		
	\end{abstract}
	

\section{Introduction}
\textrm{Disruption is a sudden collapse of tokamak plasmas. The avoidance, prediction, and mitigation of disruptions are believed to be crucial to the safe steady state operation of tokamak devices and future fusion reactors. Over the past two decades, the injection of impurity gas or pellets has been proven highly effective and robust in triggering and accelerating the onset of disruption, thus has been widely adopted as a disruption mitigation scheme \cite{Gerasimov_2020,Hollmannn2019}. Whereas many attribute the early initiation of disruption in part to the impurity radiation driven tearing mode (TM) growth, the exact mechanism underlying the impurity radiation and tearing mode interaction, as well as its universal nature in connection to the density limit disruption, remains a subject of fundamental interests and continued research \cite{Rebut1985,Rutherford1985,Suttrop_1997,Salzedas2002,Xu_2017,Gates2012,Gates_2013,Gates2015a,Gates2016,White1977,White2015,Xu2020a,Teng_2018}.}

The primary mechanism proposed for the radiation-driven tearing mode growth is the thermal instability of magnetic island due to the temperature dependent resistivity change originally developed in theories for the nonlinear tearing mode \cite{Rebut1985,Rutherford1985}. Later, several experimental observations of tearing mode growth and impurity radiation in Ohmic heating tokamaks may have been able to fit well to the scenario of the thermal instability, the direct and unambiguous validation of the mechanism is still missing \cite{Suttrop_1997,Salzedas2002,Xu_2017}. The thermo-resistive TM theory was further formally refined and extended to construct a model for the Greenwald density limit \cite{Gates2012,Gates_2013,Gates2015a,Gates2016,White1977,White2015}. Recent MHD simulations have demonstrated the impurity radiation effects on the magnetic island that are consistent with the thermal-resistive TM theory \cite{Xu2020a,Teng_2018}.
However, those simulation findings rely on the pre-existing small islands that initially grow from the linear tearing unstable equilibrium profile where the tearing mode asymptotic matching parameter $\Delta'>0$. It is not clear whether the satisfaction of the onset criterion alone, i.e. the radiation power loss exceeds the Ohmic heating power would be sufficient for the tearing mode growth in absence of a pre-existing island when the equilibrium is initially stable to linear tearing with $\Delta'<0$.

In tokamaks, a finite plasma $\beta$ is known to introduce into $\Delta'$ a resistive interchange correction involving the $D_R$ parameter, which is in general negative and stabilizing to the linear tearing mode \cite{Glasser1975,Glasser1976}. Nonlinearly, the resistive interchange effects can give rise to a perturbed Pfirsch-Schl\"{u}ter (PS) current that contributes to the development and saturation of the tearing mode island \cite{Kotschenreuther1985,Hegna99,Lutjens2001}. These resistive interchange effects, however, may bring in a different mechanism for the radiation-driven tearing mode growth in presence of a localized impurity deposition or concentration. In particular, as found in our recent simulations, the impurity distribution around a rational surface as a results of either injection or accumulation can lead to a localized radiative cooling and the reversal of $D_R$ from negative to positive. Such a $D_R$ reversal enables the conventionally stabilizing resistive interchange effects to directly drive the onset of the tearing mode linearly. And the continuous radiation cooling in the resistive layer results in the variation of the plasma resistivity, which mainly drives the nonlinear island growth till saturation against the stabilizing effect from perturbed PS current. It is worth mentioning that the $D_R$ reversal and the destabilizing effects of resistive interchange has been known to take place in the negative central shear configuration and lead to bursting MHD activity in a different context and scenario \cite{Chu1996}.
The tearing mode triggering through $D_R$ reversal in a tokamak may also provide a new avenue to the long-standing problem of seeding mechanism for the neoclassical tearing mode in high-$\beta$ tokamak plasma regime (e.g. Ref \cite{Sauter1997}).

In this paper, we report our initial findings that demonstrate the tearing mode growth triggered by the $D_R$ reversal due to a localized impurity radiation cooling around the rational surface in a tokamak. The reversal of $D_R$ leads to a positive $\Delta'$ for the linear tearing instability, and a nonlinear perturbed PS current for the finite width island growth. In addition, thermal conductivity $\kappa_{\parallel}$ affects the mode growth through $D_R$ by fast parallel thermal equilibration. These findings may suggest an additional and novel mechanism for the impurity radiation driven growth of tearing mode and magnetic island.

\section{Simulation model and setup}
Our simulations are based on the 3D resistive-MHD model implemented in the NIMROD code \cite{Sovinec2004} with an impurity radiation module KPRAD \cite{KPRAD}, and the combined system of equations are as follows	
\begin{eqnarray}
	\rho \frac{d \vec{V}}{dt} = - \nabla p + \vec{J} \times \vec{B} + \nabla \cdot (\rho \nu \nabla \vec{V})
	\label{eq:momentum}
	\\
	\frac{d n_i}{d t} + n_i \nabla \cdot \vec{V} = \nabla \cdot (D \nabla n_i) + S_{ion/3-body}
	\label{eq:contiune2}
	\\
	\frac{d n_{Z,Z=0-10}}{d t} + n_Z \nabla \cdot \vec{V} = \nabla \cdot (D \nabla n_Z) + S_{ion/rec}
	\label{eq:contiune3}
	\\
	n_e \frac{d T_e}{d t} = (\gamma - 1)[n_e T_e \nabla \cdot \vec{V} + \nabla \cdot \vec{q_e} - Q_{loss}]
	\label{eq:temperature}
	\\
	\vec{q}_e = -n_e[\kappa_{\parallel} \hat{b} \hat{b} + \kappa_{\perp} (\mathcal{I} - \hat{b} \hat{b})] \cdot \nabla T_e
	\label{eq:heat_flux}
	\\
	\frac{\partial \vec{B}}{\partial t} = \nabla \times \left( \vec{V} \times \vec{B} \right) - \nabla \times \left(  \eta \vec{j} \right) 
	\label{eq:ohm}
\end{eqnarray}	
Here, $n_i$, $n_e$, and $n_Z$ are the main ion, electron, and impurity ion number density respectively, $\rho$, $\vec{V}$, $\vec{J}$, and $p$ the plasma mass density, velocity, current density, and pressure respectively, $T_e$ and $\vec{q}_e$ the electron temperature and heat flux respectively, $D$, $\nu$, $\eta$, and $\kappa_{\parallel} (\kappa_{\perp})$ the plasma diffusivity, kinematic viscosity, resistivity, and parallel (perpendicular) thermal conductivity respectively, $\gamma$ the adiabatic index, $S_{ion/rec}$ the density source from ionization and recombination, $S_{ion/3-body}$ also includes contribution from 3-body recombination, $Q_{loss}$ the energy loss, $\vec{E} (\vec{B})$ the electric (magnetic) field, $\hat{b}=\vec{B}/B$, and $\mathcal{I}$ the unit dyadic tensor.
The impurity radiation includes bremsstrahlung, ionization, recombination, and line radiation, which all contribute to the radiation power loss $Q_{loss}$, and the line radiation ($\propto L_{Z(T_e)}n_e n_Z$, where $L_{Z(T_e)}$ is the line radiation coefficient) predominates in our simulations.
More details can be found in Appendix \ref{appendix}.

A stable EAST L-mode like equilibrium with an initial Neon impurity deposition is adopted in the simulation.
The initial Neon density distribution is localized on the $q=2$ rational surface, i.e. $N_{imp} = n_{imp} \exp(-((r-r_s)/L_1)^2) \exp(-\phi^2) (1+\cos\theta)$, where $n_{imp}$ is the peak impurity number density, $r_s,\theta$, $\phi$ the $q=2$ radial location, the poloidal and toroidal angles, respectively, and $L_1 = 1\ cm$ is the radial distribution width, which is used as a simple model for pellet injection deposition.
Although the peak level of local impurity deposition around the $q=2$ surface in simulation is close to the local density of main ion species, the overall line-averaged impurity level is about $1\% \sim 10\%$ of the background plasma density.  Such an impurity level in line-average is set slightly higher than the typical experimental value merely for the purpose of accelerating the dynamic process in the simulation that can be afforded by the available computational resource.
Initially the impurity is stationary on the rational surface in absence of equilibrium flow, i.e. $V_0 = 0 m/s$, and its subsequent diffusion and convection is governed by Eq. (\ref{eq:contiune3}). All other key parameters set up for the simulation are mainly based on the EAST like L-mode equilibrium (Table \ref{input}).

	\begin{table}
	\caption{\label{input} Key parameters in the simulation}
	\footnotesize
	\begin{tabular}{@{}llll}
		\br
		Parameter & Symbol & Value & Unit \\
		\mr
		Minor radius & $a$ & $0.45$ & $m$ \\
		Major radius & $R_0$ & $1.85$ & $m$ \\
		Plasma current & $I_p$ & $0.38$ & MA \\
		Toroidal magnetic field & $B_{t0}$ & $2.267$ & T\\
		Core value of safety factor & $q_0$ & $1.484$ & dimensionless \\
		Edge value of safety factor & $q_{95}$ & $9.768$ & dimensionless \\
		Core electron number density & $n_{e,core}$ & $2.3 \times 10^{19}$ & $m^{-3}$ \\
		Core electron temperature & $T_{e,core}$ & $2.959$ & $keV$\\
		Edge electron temperature & $T_{e,edge}$ & $3.875$ & $eV$\\
		Equilibrium velocity & $V_0$ & $0$ & $m/s$\\
		The core Lundquist number & $S_0$ & $6.413 \times 10^9$  & dimensionless \\
		The core resistivity & $\eta_0$ & $6.334 \times 10^{-11}$  & $\Omega \cdot m$ \\
		Perpendicular thermal conductivity & $\kappa_{\perp}$ & $1$ & $m^2/s$ \\
		Parallel thermal conductivity & $\kappa_{\parallel}$ & $10^{10}$ & $m^2/s$ \\
		Diffusivity & $D$ & $0.1$ & $m^2/s$\\
		\br
	\end{tabular}\\
\end{table}

We use $70 \times 64$ finite elements with the third order Lagrange polynomial basis functions in the poloidal plane, and two Fourier modes with toroidal numbers $n=0-1$ are considered in the toroidal direction. Simulations with toroidal modes $n=0-6$ indicate that the effects from higher $n$ modes are negligible on the main conclusion, confirming the numerical convergence of the simulation results from $n=0-1$ toroidal modes only. The tokamak plasma region in simulation is surrounded by a perfectly conducting wall without a vacuum region.
A single temperature $T_e$ is shared among different species based on the assumption of instantaneous thermal equilibration. Constant thermal conductivity is adopted in all simulations for simplicity, which agree with the more sophisticated thermal model $\kappa_{\perp} \propto T^{-1/2}B^{-2}$ and $\kappa_{\parallel} \propto T^{5/2}$ in terms of the main conclusion. In addition, the Spitzer resistivity $\eta \propto T_e^{-3/2}$ is applied to the simulations.

\section{Threshold in impurity level for TM growth}
\begin{figure}[ht]
	\begin{center}
		\includegraphics[width=0.8\textwidth,height=0.45\textheight]{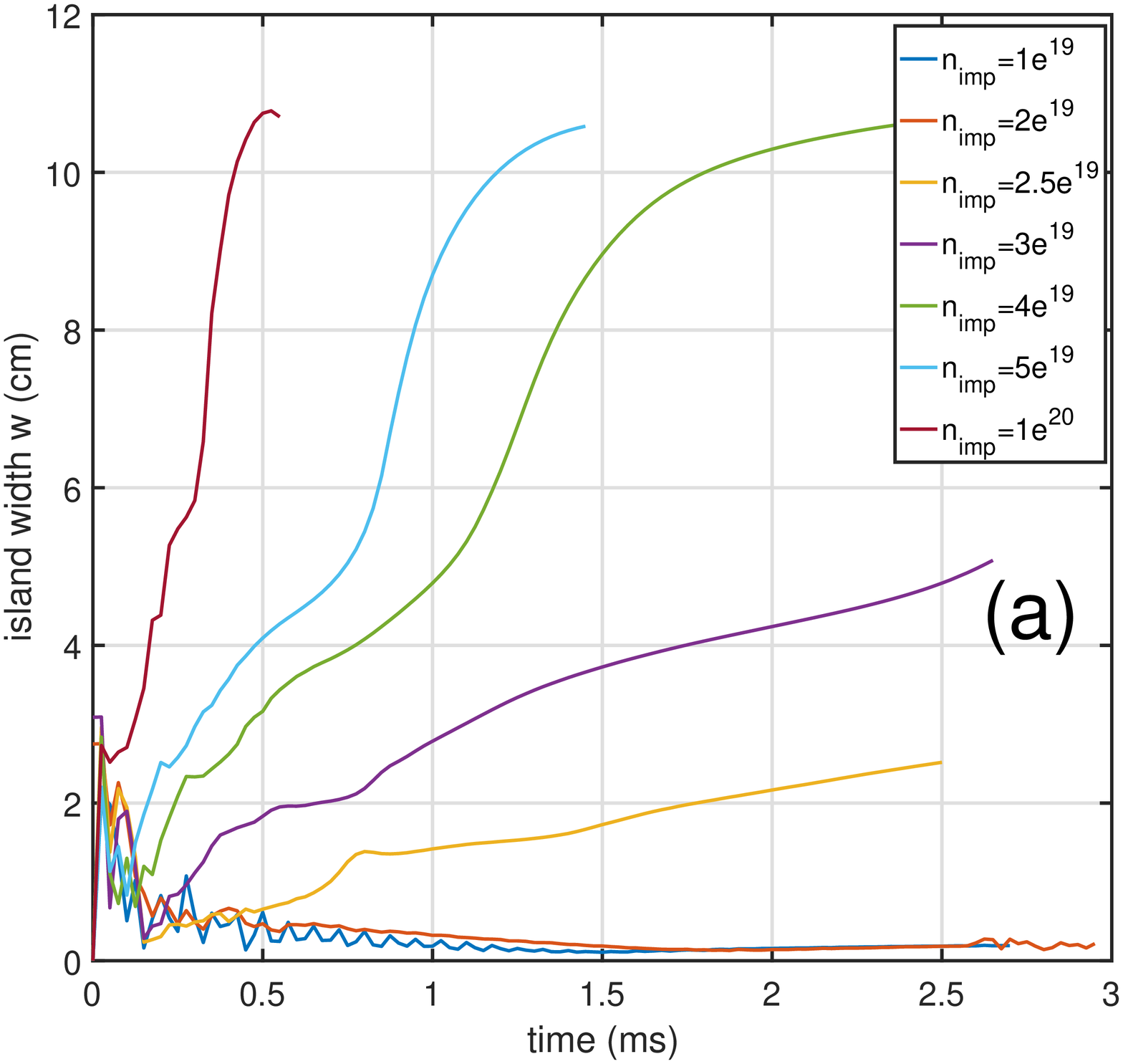}
		\includegraphics[width=0.8\textwidth,height=0.45\textheight]{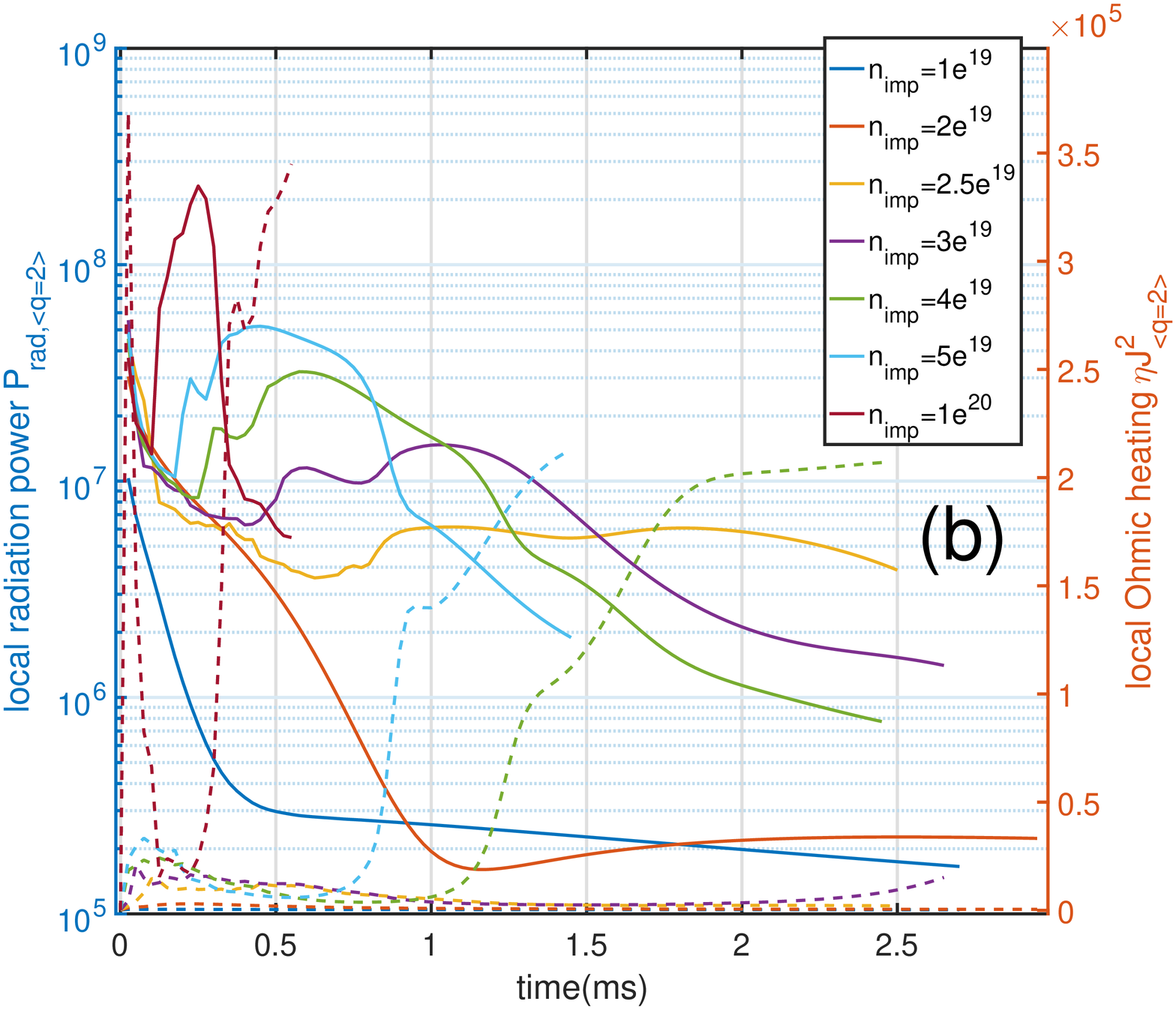}
	\end{center}
	\caption{(a) Island width, and (b) local radiation power (solid lines) and Ohmic heating (dashed lines) on the $q=2$ rational surface as a function of time for different impurity levels $n_{imp}$.}
	\label{TM-imp}
\end{figure}

In the first set of simulations, the initial level of impurity deposition is varied, and all other parameters are fixed. A threshold in the impurity level for triggering the TM growth is found, i.e. $n_{imp}^{crit}=2\times 10^{19} m^{-3}$. Below this level, the magnetic perturbation initially induced by the impurity decays away gradually and no magnetic island forms. Above this level, the island width increases till to saturation (Fig. \ref{TM-imp}a). 
However, for any initial impurity level below or above $n_{imp}^{crit}$, we found that the local radiation power largely exceeds the Ohmic heating on the $q=2$ rational surface throughout the simulation (Fig. \ref{TM-imp}b).
Thus, the criterion for the onset of the thermo-resistive instability, that the local radiation power loss exceeds the Ohmic heating power within the island \cite{Gates2012,White2015}, may not be able to explain the existence of the critical impurity level $n_{imp}^{crit}$ for the onset of the TM observed in these simulations.

\begin{figure}[ht]
	\begin{center}
		\includegraphics[width=0.8\textwidth,height=0.42\textheight]{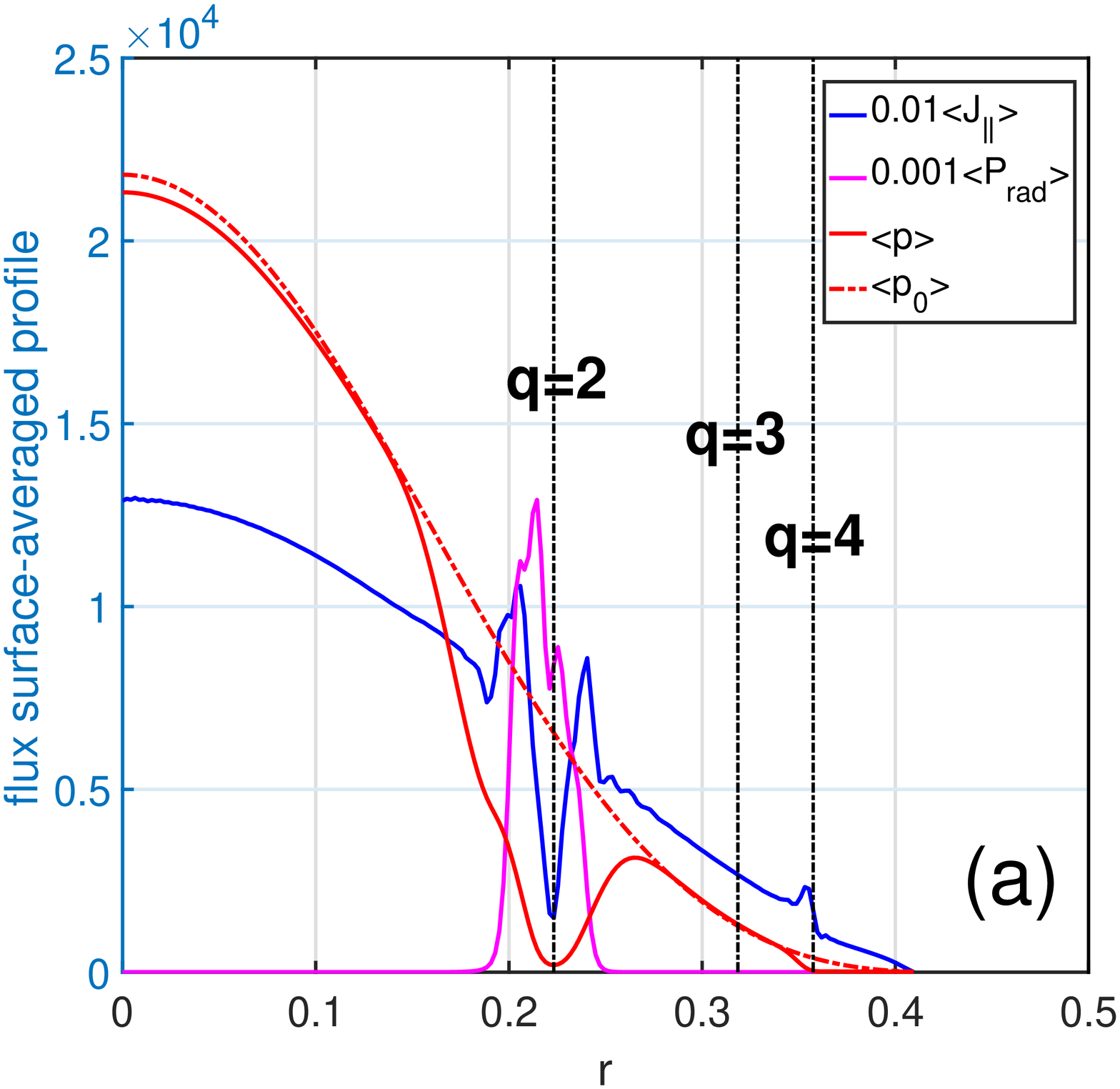}
		\includegraphics[width=0.8\textwidth,height=0.42\textheight]{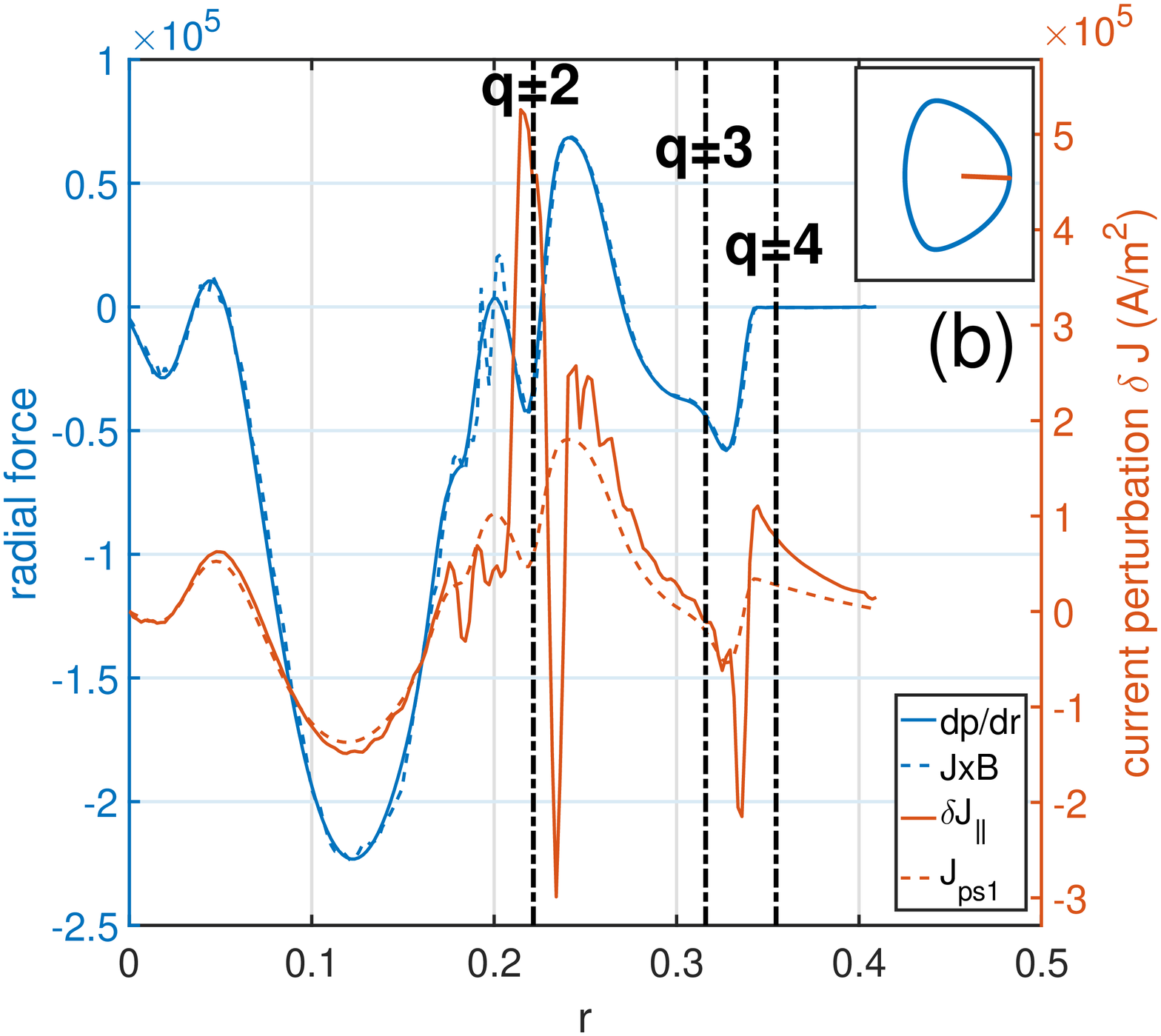}
	\end{center}
	\caption{(a) Flux surface-averaged profile for the $n=0$ component of parallel current density $J_{\parallel}$, radiation power $P_{rad}$, and pressure $p$, here $p_0$ refers to the equilibrium ; (b) Radial profile along the mid-plane on the low field side (denoted as orange line in the sketch) for the radial pressure gradient $dp/dr$, Lorentz force along the radial direction $J \times B$, the parallel current density perturbation $\delta J_{\parallel}$, and the perturbed PS current model $J_{ps1}$. The initial equilibrium $q=2,3,4$ surfaces are denoted as black dashed lines.}
	\label{force-balance}
\end{figure}

One of the most immediate consequences of the local radiation cooling from the impurity deposition on the $q=2$ surface is the formation of a narrow layer with hollow pressure profile around the same rational surface (Fig. \ref{force-balance}a). The parallel component of plasma current density varies rapidly across the layer, with a sharp dip inside and two spikes near both boundaries of the layer. To understand the nature of the parallel current profile transition around the $q=2$ surface, we note that although the initial equilibrium profiles dynamically evolve due to the influence from impurity source, the static force balance $\nabla p = \vec{J} \times \vec{B}$ remains well satisfied along the radial direction (Fig. \ref{force-balance}b), mainly because of the weak perturbed velocity i.e. $\tilde{V} \sim 10-10^2 m/s$.	Based on this quasi-static condition, the perturbed PS current may be evaluated from the model $\delta J_{ps} = J_{ps1} \cos{\theta}$, where $ J_{ps1} = -2\frac{1}{B_{\theta}} \frac{r}{R} \frac{d\delta p}{dr} $ and $\delta p = p_{(t)} - p_{(t=0)}$, which agrees well with most part of the total parallel current perturbation $\delta J_{\parallel}$, particularly in the region outside the resistive layer (Fig. \ref{force-balance}b). Here, $\delta p$, $J_{ps1}$, and $\delta J_{\parallel}$ refer to the $(m=0, n=0)$ Fourier components of perturbations.	
Besides, the $m=2,n=1$ component of helical current inside the island increases along with the island width.

\begin{figure}[ht]
	\begin{center}
		\includegraphics[width=0.8\textwidth,height=0.45\textheight]{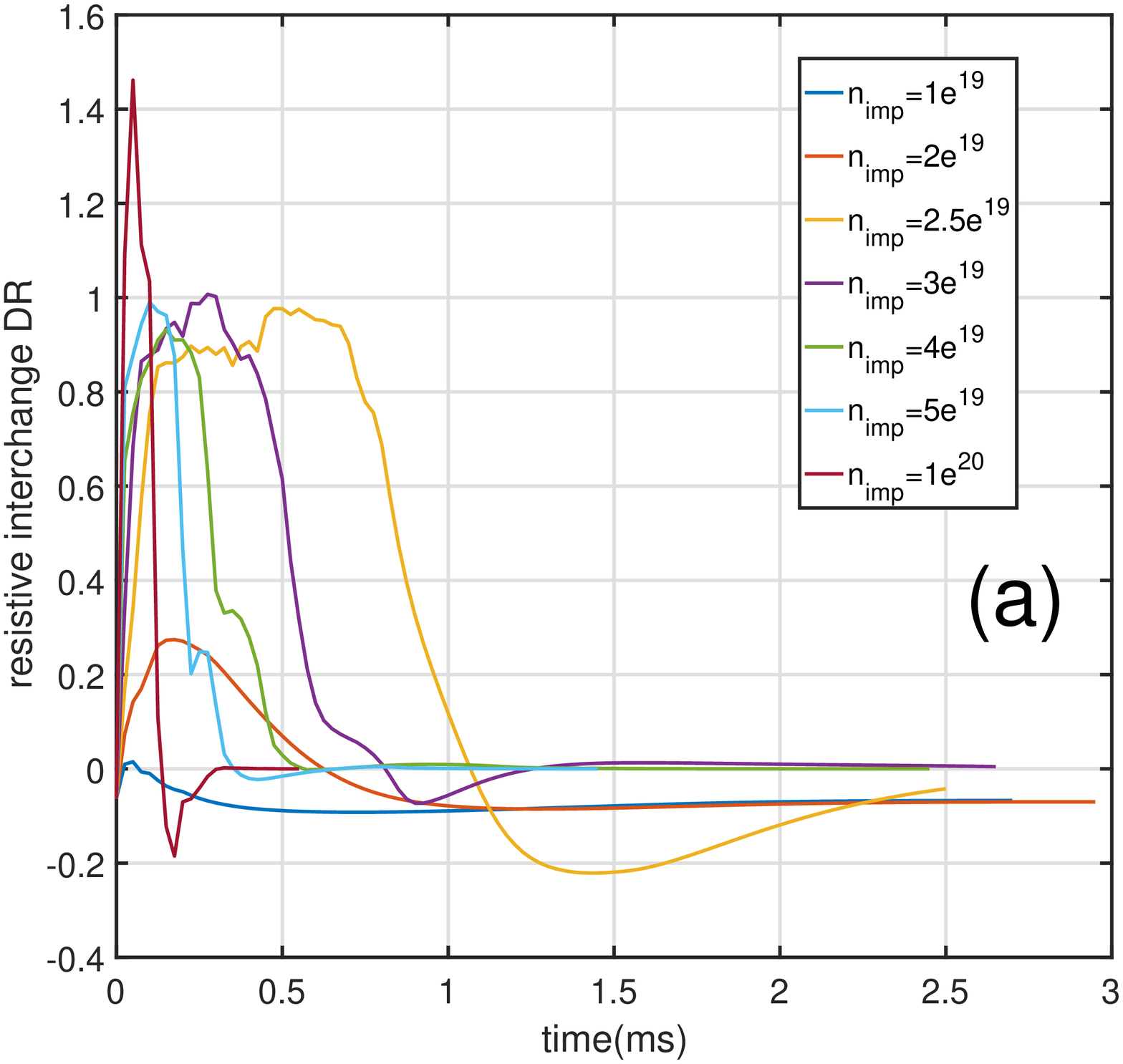}
		\includegraphics[width=0.8\textwidth,height=0.45\textheight]{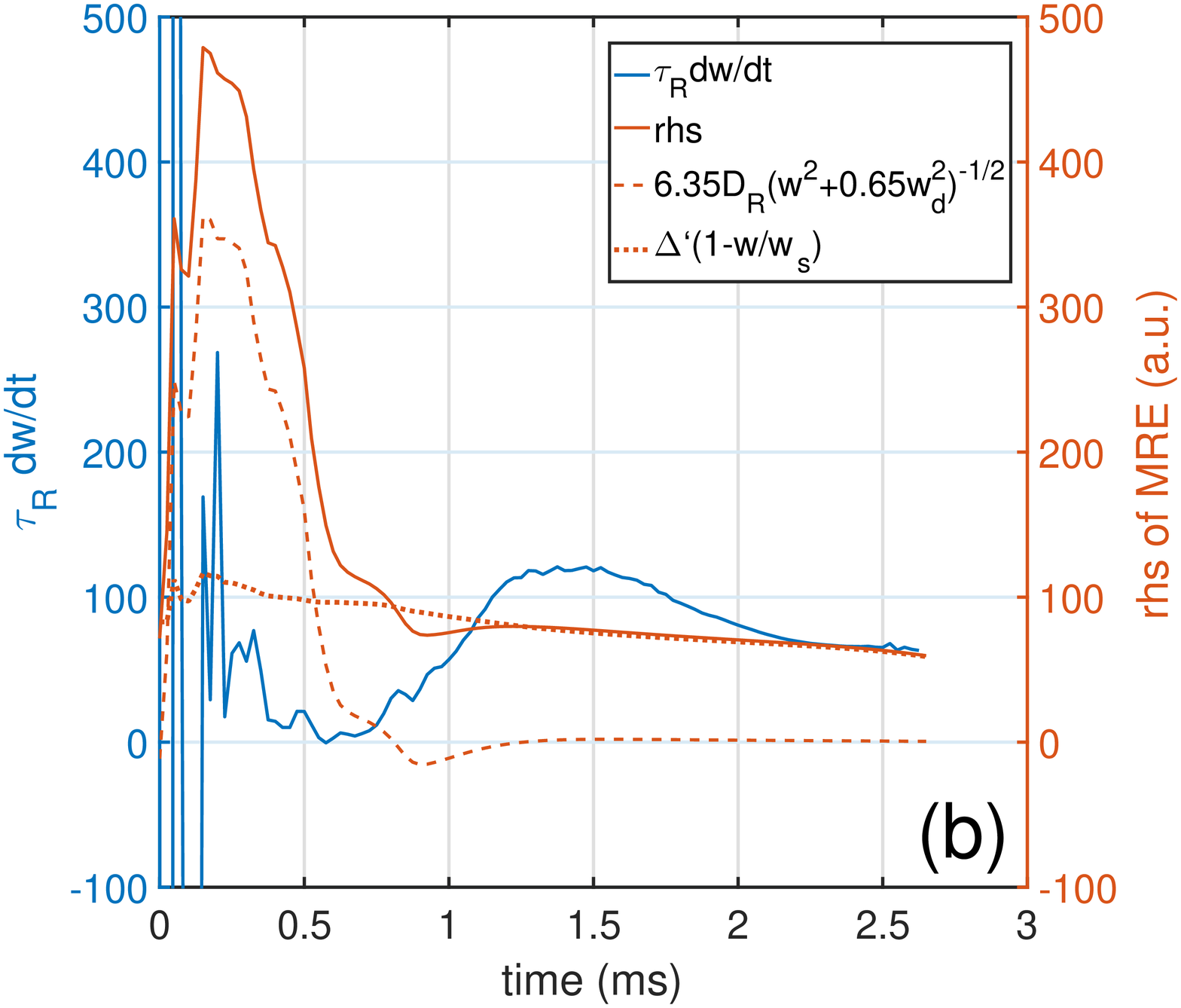}
	\end{center}
	\caption{(a) The resistive interchange parameter $D_R$, and (b) the model from Ref. \cite{Lutjens2001}, here $\Delta'=120, w_s=0.1m,$ and $w_d=0.02m$.}
	\label{delta eff discussion}
\end{figure}

The hollow pressure profile around the $q=2$ surface leads to the reversal of the resistive interchange parameter $D_R\approx \frac{\epsilon^2_s\beta_p}{s} \frac{L_q}{L_p} \left( 1-\frac{1}{q^2} \right)$, where $L_p=p/p', L_q=q/q'$, to become positive from the conventional monotonically decreasing pressure profile for which the $D_R$ is negative (Fig. \ref{delta eff discussion}a). Previous theory by L{\"{u}}tjens et al \cite{Lutjens2001} includes the $D_R$ effect on the resistive TM growth in the modified Rutherford equation
\begin{equation}
	\frac{dw}{dt} = \frac{1.22}{\tau_R}\left(\Delta'\left( 1-\frac{w}{w_s}\right) + \frac{6.35D_R}{\sqrt{w^2+0.65w_d^2}}  \right) 
\end{equation}
which predicts the destabilizing and triggering effects of the $D_R$ reversal on the TM growth. This gives the right trend of the island growth onset in simulation here, as the comparison between the above equation $(7)$ and the simulation results shows in Fig. \ref{delta eff discussion}(b) for the initial time period $t<0.5ms$. However, the absolute value of $D_R$ decreases quickly to zero and remains almost zero after $t=0.5ms$ when the island starts to grow again. This indicates that the $D_R$ reversal effect from the curvature and pressure gradient is the dominant driver in the small island region, but fails to continue supporting the island growth in the second phase after the $D_R$ effect vanish.

\begin{figure}[ht]
	\begin{center}
		\includegraphics[width=1.0\textwidth,height=0.45\textheight]{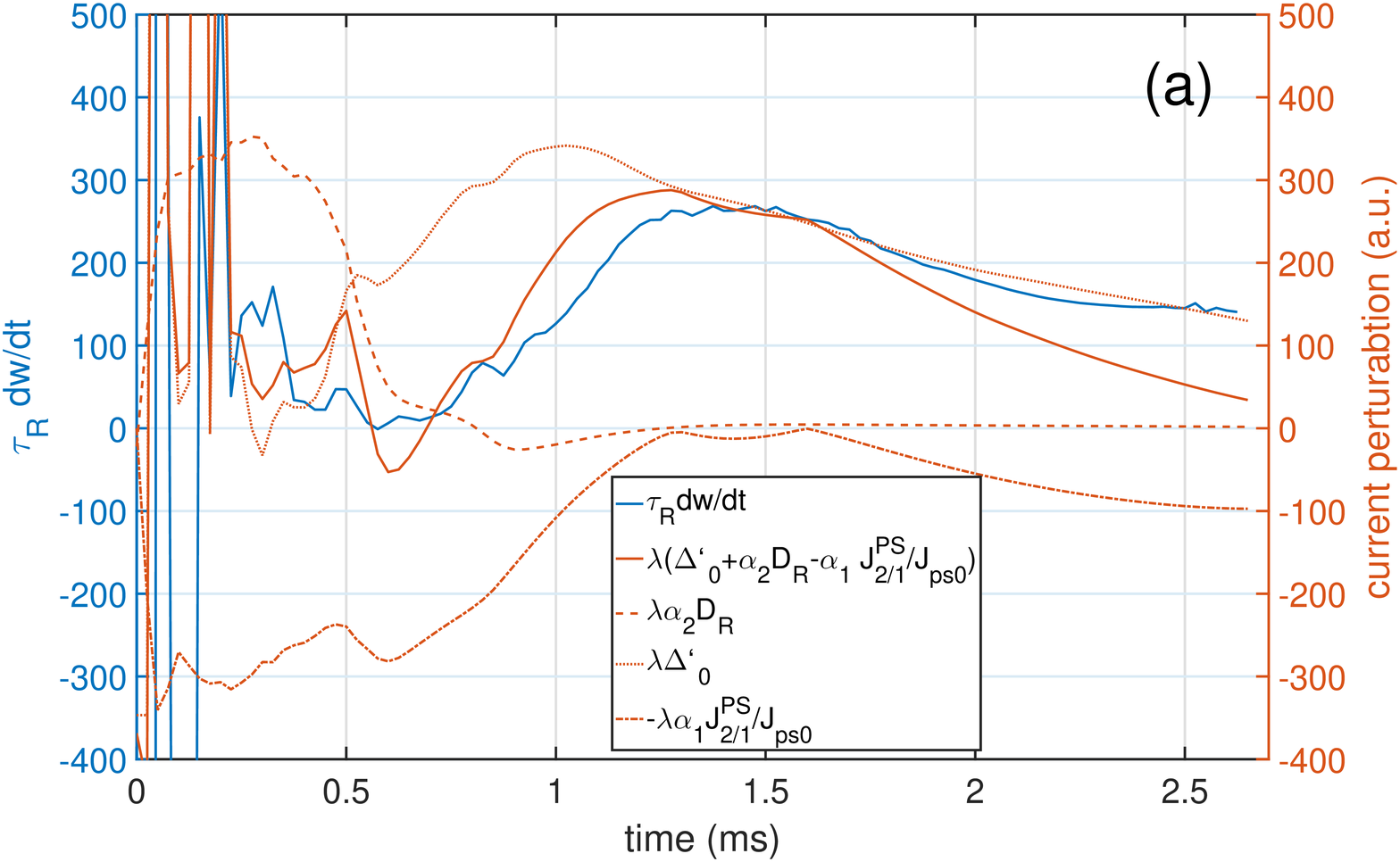}
		\includegraphics[width=0.48\textwidth,height=0.3\textheight]{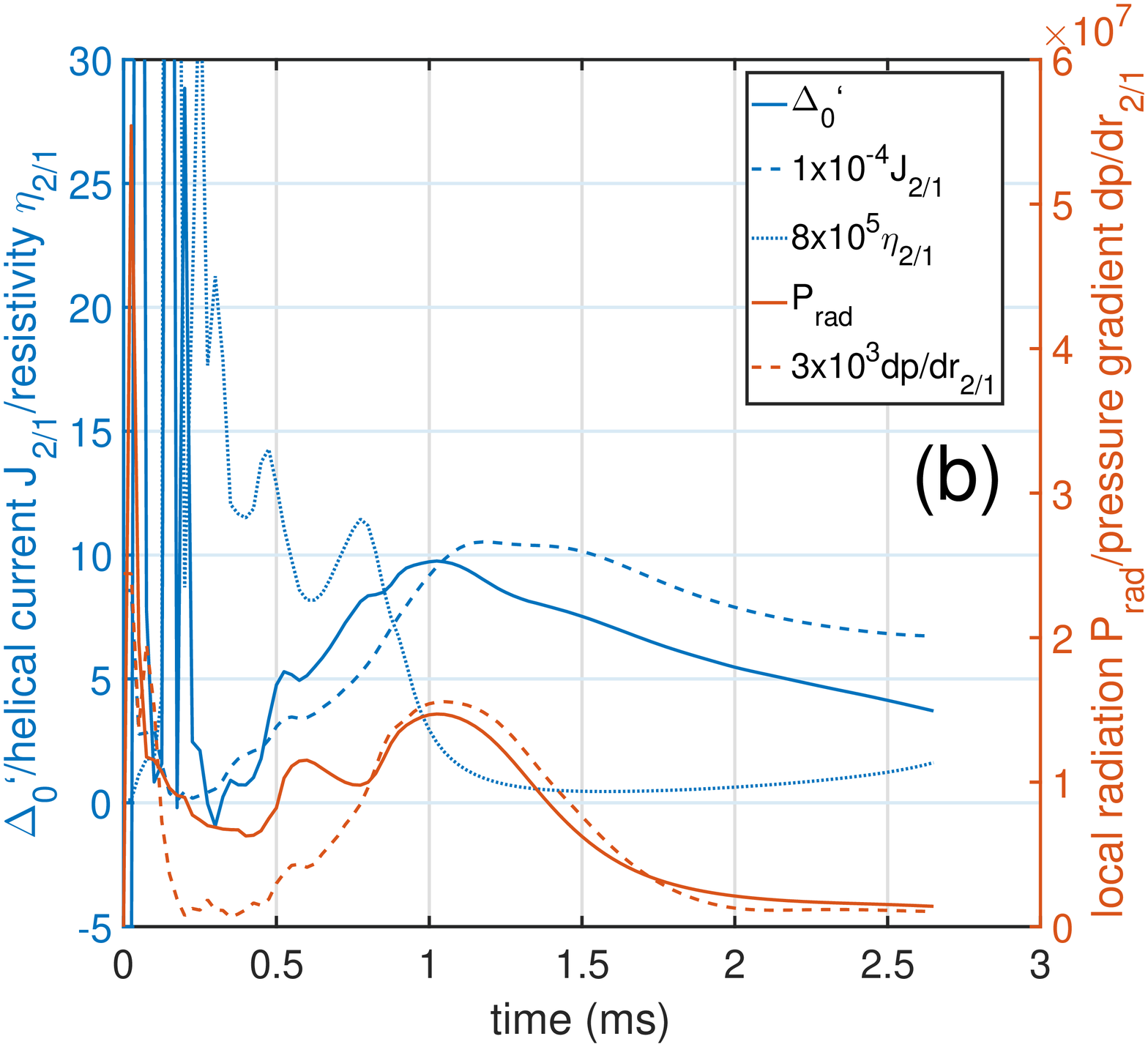}
		\includegraphics[width=0.48\textwidth,height=0.3\textheight]{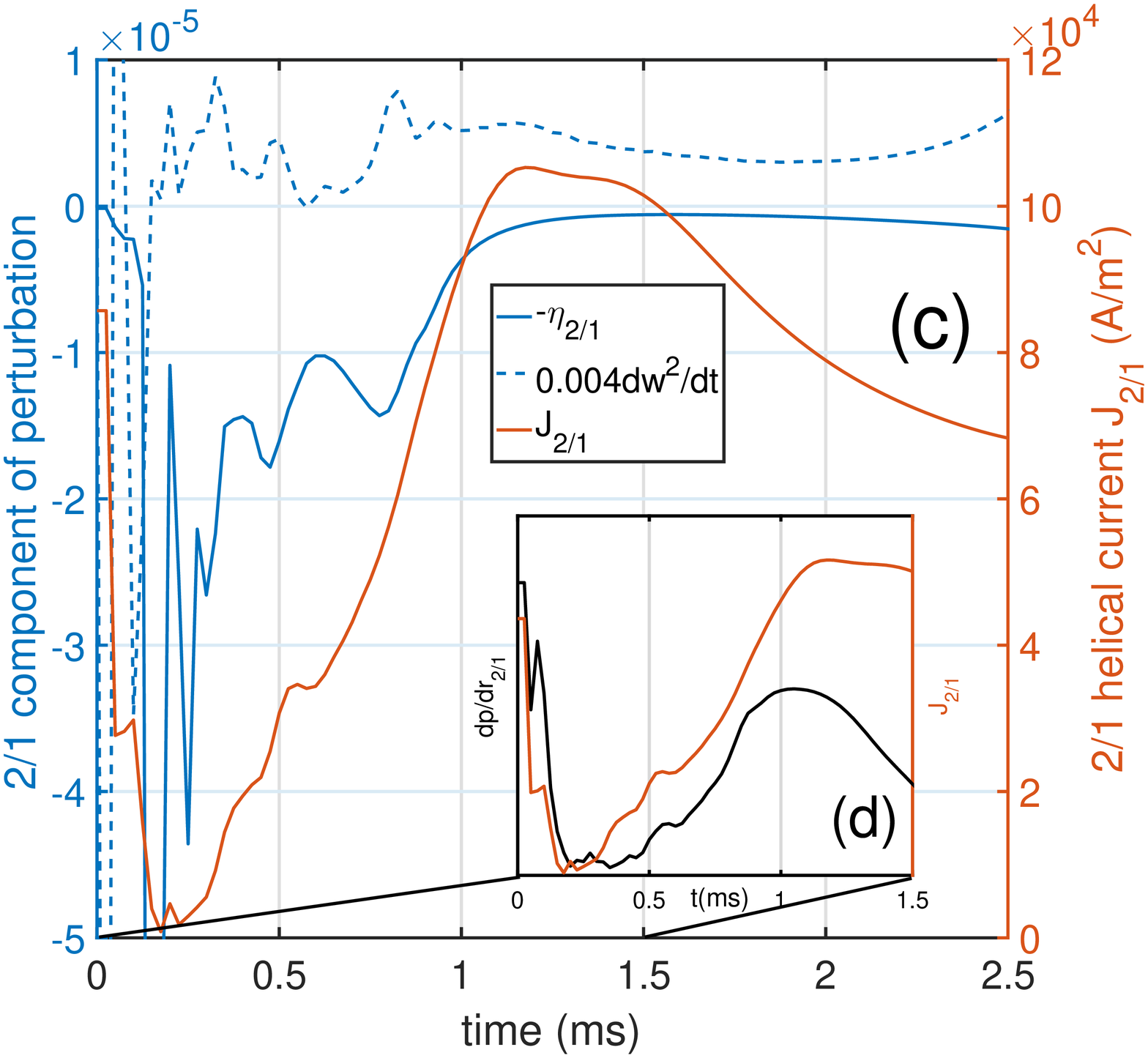}
	\end{center}
	\caption{(a) The island growth rate $\tau_R \frac{dw}{dt} $ term, the resistive interchange parameter $D_R$, the tearing stability parameter $\Delta_0'$, the $m=2/n=1$ component of PS current $J_{2/1}^{PS}$ inside the resistive layer which is normalized to the initial equilibrium PS current $J_{ps0}$, and the sum of $\Delta_0'$, $D_R$ and $J_{2/1}^{PS}/J_{ps0}$ terms, where the fitting parameters $\lambda=35,\alpha_1=0.5,\alpha_2=10$; (b) The tearing instability parameter $\Delta_0'$,  the local radiation power $P_{rad}$, the $m=2/n=1$ Fourier components of the perturbed current $J_{2/1}$, the plasma resistivity $\eta_{2/1}$, and the pressure gradient $p_{2/1}'$, respectively; (c) The plasma resistivity $-\eta_{2/1}$, the time derivative of perturbed magnetic field $dB_{r,2/1}/dt$, and the perturbed current $J_{2/1}$; (d) The perturbed pressure gradient $p_{2/1}'$ and the perturbed current $J_{2/1}$ on the $q=2$ rational surface as functions of time.}
	\label{dwdt-DR}
\end{figure}

We introduce an alternative modified Rutherford equation (MRE)
\begin{equation}
	\tau_R \frac{dw}{dt} = \lambda\left( \Delta'_0+\alpha_2D_R - \alpha_1 J_{2/1}^{PS}/J_{ps0}\right)
	\label{eq:our MRE}
\end{equation}
may be used to better fit the simulated island growth, where $\tau_R$ is the local resistive time $\tau_R=\mu_0 r_s^2/\eta_s$, $\eta_s$ the local resistivity (Fig. \ref{dwdt-DR}a). All terms are calculated from the time dependent simulation results except for the constant fitting parameters $\alpha_1, \alpha_2$ and the $\lambda$, which may account for additional quantitative geometry and plasma effects.
$\Delta_0'= \left( \frac{dB_{r(r_s^+)}}{dr} - \frac{dB_{r(r_s^-)}}{dr} \right) /B_{r(r_s)}$ is the jump in logarithmic derivative across the equilibrium $q=2$ surface, where $B_r$ is the $m=2,n=1$ radial component of perturbed magnetic field. In our simulation, the $q=2$ location barely changes throughout the time, and the flux distortion due to magnetic island formation is negligible.
The resistive interchange parameter $D_R$ includes the effect of the flux surface averaged magnetic curvature in the toroidal system, and introduces an initial linear instability to the mode due to a reversal of the $D_R$ from negative to positive, similar to the L{\"{u}}tjens theory in Eq. $(7)$.
The $m=2$ Fourier component of PS current perturbation $J_{2/1}^{PS}$, which is normalized by the the initial equilibrium PS current $J_{ps0}$, denotes the nonlinear effect of the PS current perturbation inside the resistive layer. After the $D_R$ reversal phase ($t>0.5ms$), the island growth is primarily driven by the $\Delta_0'$ term against the stabilizing effect from the nonlinear PS current $J_{2/1}^{PS}$.

The $\Delta_0'$ term is consistent with the helical current perturbation $J_{2/1}$ measured from the simulation results (Fig. \ref{dwdt-DR}b), which may be schematically represented by the following perturbative form of eq $4.4.9$ from \cite{Tokamak_wesson}
\begin{equation}
	J_{2/1} = -\mu_0 f p_{2/1}'\left(\frac{1}{B}-\frac{\left\langle 1/B_p\right\rangle }{\left\langle B^2/B_p\right\rangle } B\right) - \frac{\eta_{2/1}\left\langle E_{\phi}B_{\phi}/B_p\right\rangle }{\eta_0^2\left\langle B^2/B_p\right\rangle } B + \frac{C \left\langle \frac{dw^2}{dt} \frac{B_{\phi}}{B_p}\right\rangle }{\eta_0\left\langle B^2/B_p\right\rangle } B
\end{equation}
The first term on the right hand side refers to the PS current perturbation $J_{2/1}^{PS}$, which results from the persistent radiation cooling around the rational surface that impedes the pressure profile flattening and enhances the local pressure gradient $p_{2/1}'$.
The second and third terms account for the induced currents resulting from the variation of plasma resistivity $\eta_{2/1}$ and magnetic field perturbation $dB_{r,2/1}/dt$, respectively (Fig. \ref{dwdt-DR}c). The correlation between the local radiation power and the $\Delta_0'$ suggests that the current perturbation $J_{2/1}$ has major contributions from the pressure gradient perturbation $p_{2/1}'$ and the plasma resistivity perturbation $\eta_{2/1}$ due to temperature perturbation.
Note that there are two peaks during the evolution of the helical current perturbation $J_{2/1}$ (Fig. \ref{dwdt-DR}c), the first narrow one corresponds to the local positive pressure gradient perturbation $p_{2/1}'$ at the beginning $t=0-0.2ms$ (Fig. \ref{dwdt-DR}d), and the following dip and the second peak ($t>0.2ms$) are mainly caused by the variation of the plasma resistivity $\eta_{2/1}$, which leads to the nonlinear growth of the magnetic island.

\section{The effect of thermal conductivity on TM growth}
\begin{figure}[ht]
	\begin{center}
		\includegraphics[width=0.8\textwidth,height=0.45\textheight]{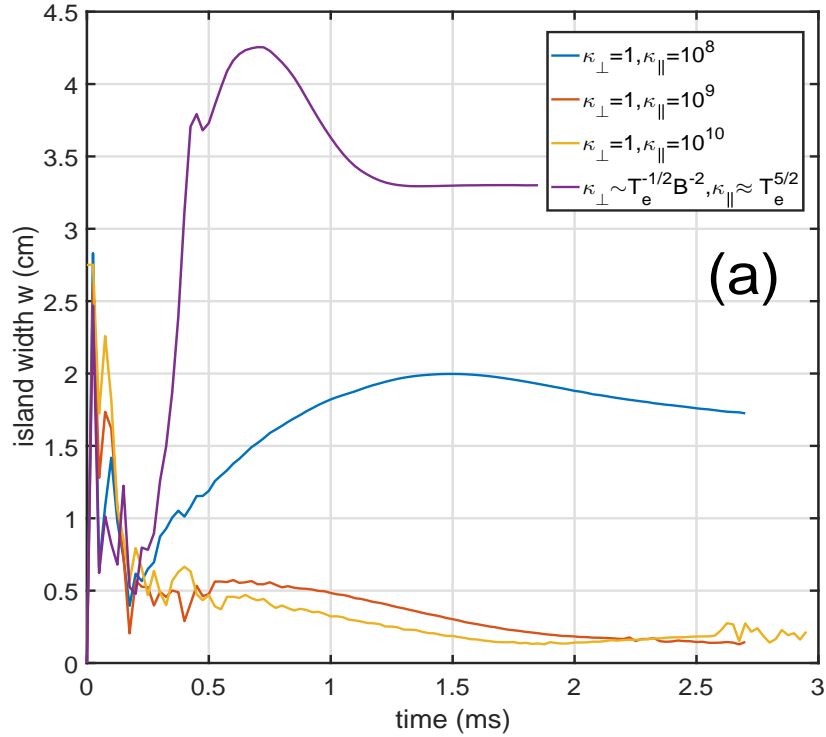}
		\includegraphics[width=0.8\textwidth,height=0.45\textheight]{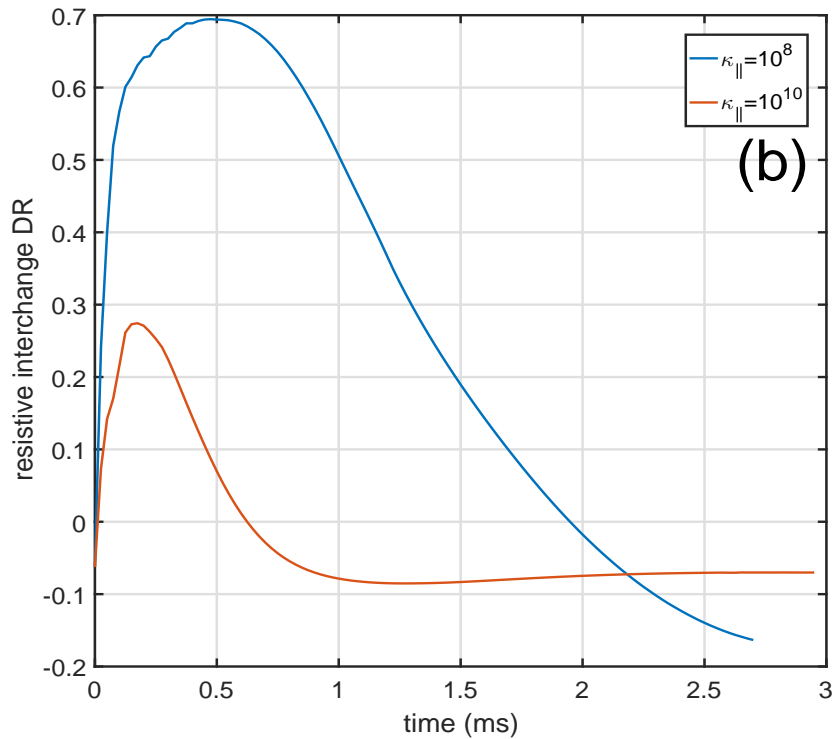}
	\end{center}
	\caption{(a) Island width as a function of time for different parallel thermal conductivity $\kappa_{\parallel}$ cases, a more sophisticated thermal model is included as well, which $\kappa_{\perp}=(3000/T_e)^{1/2}B^{-2}, \kappa_{\parallel}=10^{10}(T_e/3000)^{5/2}$. (b) Resistive interchange parameter $D_R$ as a function of time for different $\kappa_{\parallel}$ cases.}
	\label{TM-kappa}
\end{figure}

We perform another series of simulations for a range of parallel thermal conductivity $\kappa_{\parallel}=10^8-10^{10}m^2/s$ with all other parameters fixed. The impurity level is set to be the threshold found in previous section $n_{imp}=2 \times 10^{19} m^{-3}$. When $\kappa_{\parallel}$ is below $10^9 m^2/s$, the magnetic island grows up, whereas on the contrary, when $\kappa_{\parallel}$ above this value, the island saturates at a much smaller size (Fig. \ref{TM-kappa}a), which agrees with previous findings \cite{Xu2019}. In addition, simulations based on the temperature-dependent thermal conductivity model, i.e. $\kappa_{\perp} \sim T_e^{-1/2} B^{-2}, \kappa_{\parallel} \sim T_e^{5/2}$ show similar effects from the reduced $\kappa_{\parallel}$ due to radiation cooling as well.


More importantly, comparing the $\kappa_{\parallel}=10^{10}m^2/s$ case of island suppression with the $\kappa_{\parallel}=10^8 m^2/s$ case of island growth in Fig. \ref{TM-kappa}(b), the resistive interchange parameter $D_R$ in $\kappa_{\parallel}=10^8 m^2/s$ case is larger and stays positive longer than the $\kappa_{\parallel}=10^{10}m^2/s$ case.
Finite parallel thermal conductivity limits the perturbed perpendicular pressure gradient that can develop in the vicinity of the island \cite{RichardFitzpatrick1995}, and larger parallel thermal conductivity $\kappa_{\parallel}$ contributes to faster thermal equilibration along the field lines, thus weakens the radial gradient of temperature $T_e$ perturbation and allows the pressure profile to recover its initial negative gradient more quickly. Consequently the $D_R$ value switches back to negative and stabilizes the mode. These results agree with previous findings that smaller $\kappa_{\parallel}/\kappa_{\perp}$ ratio leads to more unstable tearing mode \cite{L_tjens_2001}.

\section{Summary}
A resistive interchange reversal mechanism has been identified for the impurity radiation driven tearing growth in a tokamak from NIMROD/KPRAD simulations. A localized hollow structure in the pressure profile due to impurity radiative cooling gives rise to the reversal of the resistive interchange parameter $D_R$ from negative to positive around the rational surface, which determines the trigger condition of the linear tearing mode growth and is dominant until the end of the initial small island stage. Such a $D_R$ reversal mechanism also manifests itself through the effects of the parallel thermal conductivity on the linear tearing mode growth. During the nonlinear stage, the island growth is mainly driven by the enhanced resistivity due to impurity radiation cooling. Although the crude MRE we use to fit the simulation data is able to capture the main features of the impurity radiation driven tearing mode growth, it would be more helpful to have a more complete and exact form of the MRE in order to account for not only the main mechanisms revealed in this work quantitatively, but also other secondary features neglected here, such as the vortex dynamics and the asymmetry in island structure, for examples. Such a development would require substantially more involved efforts beyond the scope of this paper, and it is planned for future work.

\section{Acknowledgments}
We thank Prof. Minghai Liu and Prof. Yonghua Ding for their helpful discussions and suggestions.
We are grateful for the supports from the NIMROD team. This work was supported by the National Magnetic Confinement Fusion Program of China (Grant No. 2019YFE03050004), the National Natural Science Foundation of China (Grant Nos. 11775221 and 51821005), the Fundamental Research Funds for the Central Universities at Huazhong University of Science and Technology (Grant No. 2019kfyXJJS193), and U.S. Department of Energy (Grant Nos. DE-FG02-86ER53218 and DE-SC0018001). This research used the computing resources from the Supercomputing Center of University of Science and Technology of China. 

\section{Appendix}
\label{appendix}

The KPRAD module adopted in the NIMROD code is used to update the impurity charge state populations and calculate the power of impurity radiations \cite{KPRAD}, which include the background impurity radiation $P_{bg}$, the line radiation $P_{line}$, the bremsstrahlung $P_{brem}$, the 3-body recombination $P_{3-body}$, the ionization $P_{ion}$ and recombination $P_{rec}$, and the atomic physics data originates from ADAS database (URL https://www.adas.ac.uk/) \cite{ADAS}.

The background impurity radiation power results from the material sputtering from the divertor or the first wall, which can be set to be beryllium $(Be)$, boron $(B)$, or carbon $(C)$
\begin{eqnarray}
	P_{bg} = f_{z,bg} \times 10^{-13} \times n_e[m^{-3}] \times 10^{p_{bg}}  \\
	p_{bg} = \sum_{i} f_{bg(i)} \times \left(\log_{10}{T_e[keV]} \right)^{i-1}
\end{eqnarray}
where $f_{z,bg} = n_{Z,bg}/n_i$ is the fraction of background impurity density, $n_i$ the plasma ion density, $n_e$ the electron density, $n_{Z,bg}$ the background impurity density, $p_{bg}$ is the polynomial as a function of the electron temperature $T_e$ based on the coronal equilibrium and $f_{bg(i)}$ are the fitted coefficients of the radiation curve, in particular, the range of temperature $T_e$ only includes $2\sim20\ keV$ in the background impurity radiation \cite{KPRAD,ADAS}.

The impurity line radiation power can be set to be helium $(He)$, beryllium $(Be)$, carbon $(C)$, neon $(Ne)$, or argon $(Ar)$
\begin{eqnarray}
	P_{line,Z(c)} = 10^{p_{line}} \times 10^{-13} \times n_e[cm^{-3}] \times n_{Z,(c-1)}[cm^{-3}] \\ 
	p_{line} = \sum_{i} f_{line(i)} \times \left(\log_{10}{T_e[keV]} \right)^{i-1}
\end{eqnarray}
where $Z$ is the atomic number and $c=0-Z$ is the impurity charge state, the line radiation of impurity charge state $c$ correlates to the density of its former charge state $n_{Z,(c-1)}$, $p_{line}$ is the polynomial as a function of $T_e$ and $f_{line(i)}$ are the fitted coefficients of the line radiation curve \cite{KPRAD,ADAS}.

The bremsstrahlung radiation power
\begin{eqnarray}
	P_{brem} = 1.69 \times 16^{-32} \times n_e^2[cm^{-3}] \times \sqrt{T_e[eV]} \times Z_{eff}  \\
	Z_{eff} = 1 + \sum_{c}^{Z} \frac{(c^2-c) \times n_c[cm^{-3}]}{n_e [cm^{-3}]}
\end{eqnarray}
where $Z_{eff}$ is the effective charge state number and $n_c$ is the impurity density of different charge state.

The 3-body recombination radiation power
\begin{equation}
	P_{3-body,Z(c)} = 8.75 \times 10^{-39} \times n_e^2[cm^{-3}] \times c^3 \times T_e^{-4.5} [eV]
\end{equation}
which is proportional to $T_e^{-4.5}$ and becomes important only at $T_e \simeq 1 eV$.

The ionization radiation power
\begin{eqnarray}
	P_{ion,Z(c)} = 1.6\times 10^{-19} \times R_{ion,Z(c)} \times n_{Z,(c-1)}[cm^{-3}] \times E_{ion,Z(c)}[eV] \\
	R_{ion,Z(c)} = n_e[cm^{-3}] \times 10^{\sum_{i} f_{ion(i)} \times\left( \log_{10}{T_e[eV]} \right)^{i-1}}
\end{eqnarray}
and the recombination radiation power
\begin{eqnarray}
	P_{rec,Z(c)} = 1.6 \times 10^{-19} \times R_{rec,Z(c)} \times n_{Z,(c)}[cm^{-3}] \times\left( E_{ion,Z(c)}[eV] + T_e \right) \\
	R_{rec,Z(c)} = 5.2 \times 10^{-14} \times n_e[cm^{-3}] \times \left( c+1\right) \times \sqrt{\frac{E_{ion,Z(c)}[eV]}{T_e[eV]}} \times f_{rec} \\
	f_{rec} = 0.43 + 0.5 \times \log_{10}{\frac{E_{ion,Z(c)}[eV]}{T_e[eV]}} + 0.469\times\left(\frac{E_{ion,Z(c)}[eV]}{T_e[eV]} \right)^{-1/3}
\end{eqnarray}
where $E_{ion,Z(c)}$ is the ionization energy of impurity charge state $c$, $R_{ion,Z(c)}$ is the ionization rate and $f_{ion(i)}$ are the fitted coefficients of the polynomial for the ionization radiation curve, $R_{rec,Z(c)}$ is the recombination rate as a function of electron density $n_e$ and temperature $T_e$ \cite{KPRAD,ADAS}. The ionization is closely associated with the recombination and note that the ionization radiation of impurity charge state $c$ correlates to the density of its former charge state $n_{Z,(c-1)}$. Besides, the ionization and the recombination rates $R_{ion,Z(c)}$ and $R_{rec,Z(c)}$ are used to update each impurity charge state density respectively in the source terms $S_{ion}$ and $S_{rec}$ of the continuity equation at every time step.

\section{Reference}
\bibliographystyle{iopart-num}
\bibliography{easttm}


\newpage

\newpage

\newpage

\newpage

\newpage


\end{document}